\documentstyle[preprint,eqsecnum,aps,floats]{revtex}
\begin{document}
\draft
\tighten
\preprint{UCSBTH-96-15, NSF-ITP-96-59, gr-qc/9703021}
\title{Comparing Formulations of Generalized Quantum Mechanics
for Reparametrization-Invariant Systems}
\author{James B.~Hartle\thanks{e-mail address:
hartle@itp.ucsb.edu}}
\address{Institute of Theoretical Physics \\ University of California
\\ Santa Barbara, CA 93106 }

\author{and}
\author{Donald Marolf\thanks{
e-mail address: marolf@suhep.phy.syr.edu}}
\address{201 Physics Building}
\vskip .26 in
\address{Syracuse University,
Syracuse, NY 13244}
\date{\today}
\maketitle

\begin{abstract}
A class of decoherence schemes is described for implementing the
principles of generalized quantum theory in reparametrization-invariant
`hyperbolic' models such as minisuperspace quantum cosmology.
The connection with
sum-over-histories constructions is exhibited and the physical
equivalence or inequivalence of different such schemes is
analyzed.  The discussion focuses
on comparing constructions based on the
Klein-Gordon product with those based on the induced (a.k.a. 
Rieffel, Refined Algebraic, Group Averaging, or Spectral Analysis)
inner product.  It is shown that the Klein-Gordon
and induced products can be simply related for the models of interest.
This fact is then used to establish isomorphisms between certain
decoherence schemes based on these products.
 
\end{abstract}


\vfill\eject

\section{Introduction}
\label{sec:I}

Generalized quantum mechanics is a comprehensive framework for 
quantum theories of closed systems. 
 \cite{Har91a,Harpp,Ishamsum}. This framework includes the usual quantum
generalizations that may be necessary for
quantum theories of dynamical spacetime geometry. Indeed, the framework
permits many such
generalizations corresponding to different ways in which 
its principles can be  
implemented. 

Time reparametrization invariance is a characteristic feature, both
classically and quantum mechanically,  of
dynamical theories of spacetime such as general relativity.  
This paper is concerned with
generalized quantum theories that incorporate this invariance
for model systems with a single reparametrization
invariance. The relativistic world line or the homogeneous
minisuperspace models in quantum cosmology are simple examples.

A quantum theory of 
a reparametrization invariant system must deal with the 
constraint associated with the invariance. 
In Dirac quantization the physical states are selected from
an extended space of states by the condition that they satisfy
an operator form of the classical constraint.
In this paper we introduce a class of generalized quantum theories
that incorporate the constraints associated with reparametrization
invariance in a natural way.  They will be called ``product space 
decoherence schemes''. We also show how earlier sum-over-histories
generalized quantum theories for systems with constraints can 
be formulated as members of this class. We show how different members of the
class can be equivalent in the sense of leading to the same 
physical predictions.

The objective of a generalized quantum theory is the prediction of the
probabilities of the individual members of sets of alternative
coarse-grained histories of a closed system given the boundary
conditions that define initial and final conditions for the system.  In
the case of a model cosmology, these histories might be alternative
histories of the four-dimensional geometry of the model universe.
Only sets of
histories that have negligible interference between their members
can be consistently assigned
probabilities \cite{Gri84,Omnsum,GH90a}. A central element in a
generalized quantum theory is therefore the decoherence functional
$D(\alpha^\prime, \alpha)$ that incorporates the system's boundary
conditions and measures the interference between pairs
of histories $(\alpha^\prime, \alpha)$ in a coarse-grained set. A 
set
of histories is said to (medium) decohere when the ``off-diagonal''
elements of $D$ are negligible. The probabilities $p(\alpha)$ of
a decoherent set of histories are the
diagonal elements of $D$. These definitions are summarized by     
\begin{equation}
D\left(\alpha^\prime, \alpha\right) = \delta_{\alpha^\prime\alpha}
p(\alpha)\ .
\label{oneone}
\end{equation}
Generalized quantum theory can be defined axiomatically \cite{Ishamsum}
and various specific constructions of decoherence functionals consistent
with these axioms have been given. The most general representation is
that of Isham, Linden, and Schreckenberg \cite{ILS94}.   We
will refer to a particular {\it algorithm} for constructing
decoherence functionals as a `decoherence scheme'. 

In this paper we
consider a specific class of `product space' 
decoherence schemes for model systems with a single reparametrization 
invariance.  In these schemes, initial and final boundary conditions
are represented by physical states that satisfy the constraints
in an extended space of states.  
Reparametrization invariant histories
are represented by `class operators'
that are annihilated by the constraints.
Decoherence functionals are  
constructed by combining these elements using a product operation
$\circ$. The natural product on the extended space of 
states is typically not a candidate for $\circ$, because the 
product between two physical states 
diverges in that product. Rather,
a special construction is required for the inner product between
physical states, and this paper is concerned with the relationship
between the decoherence schemes defined with different choices
for $\circ$. In particular, we investigate
the relationship between decoherence
schemes defined through the Klein-Gordon product $\circ_{KG}$ employed,
for example, in the sum-over-histories constructions of 
\cite{Harpp,Whe94,CHup}, and those
defined through a product which
been discussed under a variety of names 
that include `spectral analysis' \cite{Marcel}, `group averaging' \cite{Higsum},
`refined algebraic quantization' \cite{ALMMT95,Mar95b} and the `Rieffel
induced' \cite{Lan95} inner product.  We shall refer to it here 
as simply the `induced product' $\circ_I$.  The definition of this
product will be reviewed in Section III.
The induced 
product is a promising route to a rigorous definition of the
inner product between physical states in non-perturbative canonical
quantum gravity (see, for example \cite{ALMMT95,Mar94,Mar95a,BDT,PI}
for results
concerning this inner product).
The relationship between the canonical and
sum-over-histories approaches to the quantum mechanics of spacetime
is of great interest. Relating the decoherence schemes based on the
natural products associated with the two approaches is a step in
the direction of clarifying that connection. 

This relationship between the Klein-Gordon and induced products in 
particular cases raises the possibility that the decoherence
schemes based on the two products could also be related or could even
be even isomorphic.  An important consideration, however, is that
decoherence schemes involve more structures than just this product.
 A precise definition of a decoherence scheme and
of an isomorphism between decoherence schemes is needed to give
meaning to this question; it will be provided in Section II.

We begin in Section II with various preliminaries and
definitions: a description of the
class of reparametrization-invariant models to be considered (the
asymptotically free hyperbolic or AFH models), a 
full definition of a decoherence scheme, a description of the
product space schemes (the schemes on which we will focus), and a 
discussion of how the sum-over-histories schemes of 
\cite{Harpp} can be written as product space decoherence schemes.
Section III contains a 
description of the product $\circ_I$ and
shows that it is related to the Klein-Gordon product
for many asymptotically free hyperbolic models.
Section IV  constructs
isomorphisms between decoherence schemes based on the two
products.  Section VII contains a
summary and some brief conclusions.

\section{Preliminaries}
\label{sec:II}

This section contains various background material for our main
discussion.  In particular, we introduce the 
notion of a decoherence scheme, describe the
class of (asymptotically free hyperbolic [AFH]) 
models to be considered, 
describe the class of `product space'
decoherence schemes,  
and finally show that the product space 
schemes described above are sufficiently general to include
the sum-over-histories constructions of \cite{Harpp}.  This sets the
stage for further discussion of the products in Section III and the
construction of isomorphisms in
Section IV.

\subsection{A Definition of Decoherence Scheme}

We shall be concerned in this work with comparing different ways of
constructing decoherence functionals for theories with a single
reparametrization invariance. We aim not merely to compare 
the decoherence functionals that might be appropriate for 
different systems, but rather to compare the different {\it algorithms}
or {\it processes} or {\it methods} of constructing decoherence functionals
for this broad class of theories, somewhat in the way that it is
possible to compare different ``methods of quantization''\footnote{
The Isham-Linden-Schreckenberg \cite{ILS94} theorem gives a standard 
representation by which to compare any two decoherence functionals.
However, here we aim at comparing different algorithms for 
constructing decoherence functionals in the specific class of
reparametrization invariant theories. For this purpose the more
restrictive notion of decoherence scheme introduced below is useful.}
In particular we shall be interested in the question of when two
different algorithms for the construction of decoherence functionals
give physically equivalent results. 

A precise but still broad notion of an algorithm for the construction 
of decoherence functionals is provided by the idea of a {\it 
decoherence scheme} which we now introduce. We begin
with the specification of a vector space ${\cal V}_N$ of `states' 
together with a space of `class operators' ${\cal C}$ acting on
${\cal V} $ which represent
the possible individual histories of system. The space  $\cal C$
contains a preferred unit operator $U$ representing the identity
alternative, and of course zero representing the empty alternative. 
`Exhaustive' sets of alternative histories $\Pi$ (for "partition'') 
are certain subsets
of $\cal C$ satisfying in particular the condition  
$\sum_{C \in \Pi} C = U$ under some appropriate addition operation. 
A {\it decoherence scheme} over $({\cal V}_N,{\cal C})$ is then just the 
assignment of a decoherence functional $D(\phi,\psi, \Pi)$ to 
each triple $(\phi, \psi, \Pi )$ where $\psi \in {\cal V}_N$ is the
`initial condition,' $\phi \in {\cal V}_N$ 
is the `final condition,' 
and $\Pi \subset {\cal C}$ is an exhaustive set of alternative histories.
(Such sets $\Pi$ will also be called `partitions of the space of
histories' or just `partitions' for short.)

The example of standard quantum mechanics may help to make these ideas
concrete. There the space ${\cal V}_N$ is the usual Hilbert space of
normalizable states. 
Partitions $\Pi$ are formed from sequences of
orthogonal projections taken from exhaustive sets, and the space $\cal C$
is the space of all the strings of projections made up from these sets.
The unit
alternative is the identity operator. 
It should be clear from this example that the space $\cal C$ 
has a considerable amount of structure arising from the conditions which
determine just what is an admissible history and the requirements
for  sets of alternative histories. 
The preferred unit operator $U$ is the most general example of such
structure. Another is the closure of $\cal C$ under addition if
the logical `or' operation holds for histories. In that case
$\cal C$ has the structure of an Abelian semi-group. 

In the above definition, 
${\cal V}_N$ is a space of `pure' initial and final conditions
(in the language of \cite{Harpp}).  Decoherence functionals
for `mixed states' can be defined in terms
of those for the pure states. 
Indeed, such mixed initial and final conditions are
necessary for realistic cosmological models where the
final boundary condition is likely to be indifference with respect to
final state (See, {\it e.g.}~\cite{GH93b}). If $I=(\{\psi_a\}, \pi^i_a)$ is a
set of initial states together with their probabilities $\pi^i_a$ and
$F=(\{\phi_b\}, \pi^f_b)$ is the corresponding set for a final condition,
then the decoherence functional for such  mixed initial and final states
 is just
\begin{equation}
D( F,I,\Pi)= {\cal N} \sum\nolimits_{ab} 
\pi^f_a \pi^i_b D(\phi_a,\psi_b,\Pi),
\label{fourtwo}
\end{equation}
the normalizing constant ${\cal N}$ being chosen so $D=1$ when ${\Pi}$
contains only the unit alternative.

A general notion of equivalence between decoherence schemes may be introduced
as follows:  Decoherence
schemes over $({\cal V}_{N,1},{\cal C}_1)$ and over $({\cal V}_{N,2},{\cal C}_2
)$
will be called  {\it isomorphic} when there exist isomorphisms
$\sigma: {\cal V}_{N,1} \rightarrow {\cal V}_{N,2}$ and $\chi: {\cal C}_1
\rightarrow {\cal C}_2$ such that 
$D_1(\psi,\phi,\Pi) = D_2(\sigma(\psi),\sigma(\phi),
\chi(\Pi))$ where $\chi(\Pi) = \{ \chi({C}): {C}\in \Pi \}$ and such
that $\chi$ places the partitions of $U_1$ and $U_2$ in
one-to-one correspondence.  When two
decoherence schemes are isomorphic, they encode the same structure; 
all of their predictions, both for decoherence and probabilities, 
are the same in the sense that any result derived for states
$\psi,\phi$ and a partition $\Pi$ in the first scheme can also be
derived for some states $\psi',\phi'$ and some partition $\Pi'$ in the second
scheme. The two schemes are then physically equivalent. 

The simplest example of an isomorphism between decoherence schemes
is provided in usual quantum mechanics by a by a unitary transformation
of the operators representing the histories and the states representing
the initial and final states \cite{GH94}. This operation clearly leaves
the decoherence functional unchanged and is an isomorphism in the sense
described above. 

The structures of the set $\cal C$ are in general altered
under an isomorphism of the kind described. For example,
the unit operator $U$ will in general be mapped to a new operator
serving the same purpose.  It is sometimes desirable to fix
these structures and, as a result, to
restrict the possible isomorphisms. For example
for reparametrization invariant systems it will prove useful
to examine isomorphisms within the class of product space decoherence
schemes (to be defined below) where the physical states and class operators
satisfy fixed constraints. In that case the mapping must preserve this
property.

\subsection{The Asymptotically Free Hyperbolic Models}
\label{subsec:II.a}

A reasonably general class of models with a single reparametrization
invariance involves a phase space
spanned by $n$ co\"ordinates $x^A$ and their conjugate momenta $p_A$, with a
(classical) constraint of the form
\begin{equation}
H_{cl}=G^{AB} p_A p_B + V(x^A)
\label{twoone}
\end{equation}
where $G^{AB}$ is a metric with signature $(-,+,\cdots,+)$ on the
configuration space ${\cal Q} = {\bf R}^n$. The coordinates $x^A$
are assumed to form a global chart on ${\cal Q}$.  We take
$({\cal Q},G)$
to be time orientable and $G$ to be asymptotically flat in the
distant past.
We frequently denote these co\"ordinates by $x$, writing for example
$V(x)$. We denote the timelike co\"ordinate $x^0$ by $t$ and the $n-1$
spacelike co\"ordinates collectively by $\vec x$. 

An obvious example is a single relativistic world line.  Then $\cal Q$ 
is four-dimensional spacetime,  the $x^A$ are
four spacetime co\"ordinates,  $G^{AB}$ is the Minkowski metric, and
$V(x^A)= m^2$ where $m$ is the particle's rest mass.
It should be emphasized, however, that in general cases $\cal Q$ is not
a spacetime even though the metric
$G$ defines a causal structure on $\cal Q$ (so that we may
use the terms `past' and `future' in this context). 
The dependence of $V$ on $x^0$ means that a 
typical classical history may wander back and forth in `time' in 
complete disregard for this causal structure.  This is quite
common in the case of cosmological models, where the `timelike' coordinate
usually describes the size or `scale factor' of the cosmology.
Thus, any classical solution that first expands and then recollapses
moves first `forwards' and then `backwards' in so-called `time.'

A more general class of examples is the set of
homogeneous minisuperspace models in quantum cosmology
\cite{Jan79,RS75,ATU93}.
Their spacetime geometries are given by metrics
of the form
\begin{equation}
ds^2= -dt^2 + e^{2\alpha} (e^{2\beta})_{ij}\ \omega^i\omega^j\ .
\label{twotwo}
\end{equation}
Here $\alpha$ is a function while
$\beta_{ij}$ is a $3\times 3$ traceless
matrix --- both depending only on $t$. The $\omega^i$ are three
one-forms whose commutation relations are the Lie algebra of a group
expressing spatial homogeneity. In the case of Bianchi IX models with a
homogeneous scalar field $\phi(t)$, we may take
\begin{equation}
x^A = \left(\alpha, \beta^+, \beta^-, \phi \right)
\label{twothree}
\end{equation}
where $\beta^\pm$ are two of the principal values of $\beta_{ij}$. The
metric  in (\ref{twoone}) is given by
\begin{equation}
G^{AB} = {\rm diag} (-1, 1, 1, 1)
\label{twofour}
\end{equation}
and the potential $V(x^A)$ by
\begin{equation}
V(x^A) = e^{4\alpha} \left[V_\beta (\beta_+, \beta_-)-1\right] + e^{6\alpha}
\left[V_\phi (\phi) + \Lambda\right]
\label{twofive}
\end{equation}
where $\Lambda$ is the cosmological constant and the functions $V_\beta$
and $V_\phi$ can be found in \cite{MTW70}.

All of these models have the feature that $V(x)$ becomes constant
as $t\equiv x^0$
tends to $-\infty$ and the metric becomes asymptotically flat
in this region:
\begin{equation}
V(x^A) \rightarrow {\rm const.}, x^0 \to -\infty \  .
\label{twosix}
\end{equation}
We shall in fact assume a slightly stronger properties in what follows.
We will be interested in models for which the metric is asymptotically
flat at past null and timelike infinity and for which the potential
becomes a nonnegative constant in this region.  While not all
of the above cosmological models fall into this class (in particular, 
the Bianchi IX potential diverges on past null infinity), models
such as Bianchi II and Bianchi V do \cite{ATU93} fulfill our requirements.
Such models will be called `asymptotically free hyperbolic models'
or AFH models for short. 

At $x^0 \to +\infty$ a variety of behaviors of $V(x^A)$ is possible. 
When the metric $G$ is asymptotically flat and
the potential $V$ approaches a constant as $x^0 \rightarrow + \infty$, 
we will say that the
model is also asymptotically free in the far future.

\subsection{The Product Space Schemes}

We consider systems with a single reparametrization invariance such as
the AFH models described above.
Recall that, as a consequence of this invariance, 
the classical canonical co\"ordinates and momenta
are subject to a constraint
\begin{equation}
H_{cl}(p_i, x^i) = 0.
\label{onetwo}
\end{equation}
For example, for the relativistic particle $H_{cl}=p^2+m^2$ where $p$ is the
particle's four-momentum.  

Construction of a product space scheme begins
by introducing a linear space of states ${\cal V}$ of smooth
`wave functions' $\psi(x^i)$ of the canonical co\"ordinates on the
configuration space ${\cal Q}$.   ${\cal V}$ should be equipped with
an involution $*$, which is usually complex conjugation of the
wave functions $\psi(x)$.  Initial and
final conditions are represented by density operators in $\cal V$ which
satisfy a quantum version of the constraints. Thus, for a pure state $\psi$,
we require
\begin{equation}
H\psi (x^i) = 0,
\label{onethree}
\end{equation}
where $H$ is an operator version of the classical constraint $H_{cl}$.
We assume that the constraint is real, 
in the sense that $\psi$ satisfies (\ref{onetwo}) if and only if
$\psi^*$ satisfies (\ref{onetwo}).
Individual coarse-grained histories (class operators)
are described by elements of
${\cal V} \otimes {\cal V}$ and are often
represented by functions (called `matrix elements')
$C(x'',x')$ on ${\cal Q} \times {\cal Q}$.
We assume that these
satisfy the constraints as well, separately in each argument.  Thus
\begin{equation}
(H \otimes \openone) C = (\openone \otimes H) C = 0.
\label{onefour}
\end{equation}
Any such operator $C$ may be chosen to be the unit alternative $U$ for the
scheme.

The last
element of the construction is a 
bilinear product operation  $\circ$.  However, this product
is not defined on the entire space ${\cal V} \times {\cal V}$.  Instead, 
it is only defined on
${\cal V}_N \times {\cal V}_N$  for some subspace ${\cal V}_N 
\subset {\cal V}$ of {\it solutions} to the constraints.  We require that
${\cal V}_N$ is
preserved under the involution $*$.
The subspace ${\cal V}_N$
will be called the space of $\circ$-normalizable states;  only ${\cal V}_N$
will provide initial and final conditions for the decoherence functionals.   
We require the product to take
values in the complex numbers and to satisfy
$(\phi\circ\psi)^*=(\psi^*\circ\phi^*)$ where $*$ denotes either 
complex conjugation or the involution on ${\cal V}$.

In a product space scheme, the decoherence functionals are constructed
in terms of the above objects as follows.
For a `pure' initial state $\phi$, a pure final state $\psi$,  and
a partition $\Pi$, the decoherence functional $D(\phi,\psi,\Pi)$ is given by
\begin{mathletters}
\label{onesix}
\begin{equation}
D(\phi,\psi,\Pi)_{\alpha^\prime, \alpha}=
{\cal N} \left(\phi^* \circ C_{\alpha'} \circ
\psi\right)\, \left(\phi^* \circ C_\alpha \circ \psi\right)^*\ .
\label{onesix a}
\end{equation}
The quantity ${\cal N}$ that ensures that the decoherence functional is
normalized is given by
\begin{equation}
{\cal N}^{-1} = |(\phi \circ U \circ \psi)|^2.
\label{onesix b}
\end{equation}
\end{mathletters}

For the typical quantum cosmology case, ${\cal V}$ is a set of functions on
the configuration space ${\cal Q}$, but 
the inner product that makes (a large subspace of) ${\cal V}$
into the Hilbert space $L^2({\cal Q})$ is not a natural choice for $\circ$ 
because none of the nontrivial solutions to the constraints
(\ref{onetwo}) are contained in this Hilbert space. For instance,
in the case of the relativistic particle,  (\ref{onetwo}) is the massive wave
equation and its solutions are not square integrable
over the whole of four-dimensional space.  At least two different
choices for $\circ$ have been proposed for a generalized quantum theory of
the relativistic world line.  The first is the standard Klein-Gordon
inner product $\circ_{KG}$ defined on a hypersurface $\sigma$ of spacetime.
\begin{equation}
\psi \circ_{KG} \phi = i\int_\sigma d \Sigma^\mu \psi^* 
\buildrel\leftrightarrow \over {\nabla}_\mu \phi.
\label{onenine}
\end{equation}
The product $\circ_{KG}$ is independent of $\sigma$ because $\psi$ and
$\phi$ satisfy the constraint.
The second  product has been referred to by many names
that include `spectral analysis \cite{Marcel}', `group averaging \cite{Higsum}',
`refined algebraic quantization \cite{ALMMT95,Mar95b} and the `Rieffel
induced \cite{Lan95}' inner product.  We shall refer to it here 
as simply the `induced product' $\circ_I$.
Very roughly, the induced product is defined by 
\begin{equation}
\int_{\cal Q} \ d x \psi^* (x) \phi (x) = 
\delta (0) \left(\psi^* \circ_I \phi\right)
\ .
\label{oneten}
\end{equation}
A more precise definition will be given in Section III below, but
for full details the reader should refer to \cite{Mar95b}
or \cite{ALMMT95,Lan95}.
Note that while the product described in these references is actually a
Hermitian inner product, we have inserted an extra application of $*$
in the definition (\ref{oneten})
in order that $\circ_I$ above is a complex bilinear
product in accordance with the structure presented above.

\subsection{Sum-Over-Histories}
\label{sec:V}

We now show how previous work on sum-over-histories
constructions of generalized quantum theories for systems with a single
reparametrization can be incorporated in the framework described in this
paper.  By generalized sum-over-histories quantum mechanics
we shall refer to the framework described in \cite{Harpp}.
Class operators are then constructed through
sums over paths of $\exp(iS)$ where $S$ is
the action for the reparametrization invariant system. A canonical
action for the systems with a single reparametrization invariance
discussed in Subsection B can be defined by introducing a multiplier
$N(\lambda)$ to enforce the constraint and writing
\begin{equation}
S\left[N(\lambda), p_A(\lambda), x^A (\lambda) \right] =
\int^1_0 d\lambda \left[p_A \dot x^A - NH\right]
\label{fiveone}
\end{equation}
where a dot denotes a derivative with respect to the parameter
$\lambda$. This action is invariant under reparametrizations
generated by $\lambda \to f(\lambda)$.

A sum-over-histories generalized quantum theory begins by positing a set of paths as
the unique set of fine-grained histories. In the present case these may
be taken to be the configuration space paths $(x^A(\lambda),
N(\lambda))$ or, slightly more generally, phase space paths
$(p_A(\lambda), x^A(\lambda), N(\lambda))$. Sets of coarse-grained
histories are partitions of these paths into reparametrization-invariant
classes $\{c_\alpha\}, \alpha=1,2,\cdots$. Each class $c_\alpha$ is an
individual coarse-grained history.

A general kind of partition of the fine-grained paths is obtained
by classifying  paths by the values
of a reparametrization-invariant functional, $F[p,x,N]$. Consider for
example the class of paths $c_\alpha$ for which $F$ lies in a range
$\Delta_\alpha$. A corresponding set of `matrix elements' would
be constructed as follows
\begin{equation}
\langle x^{\prime\prime} || \widetilde C_\alpha || x^\prime \rangle =
\int_{(x^\prime x^{\prime\prime)}} \delta p \delta x \delta N
\ e_\alpha[F] \Delta_\Phi \delta [\Phi] \exp(iS[p,x,N]).
\label{fivethree}
\end{equation}
Here $e_\alpha$ is the characteristic function for the interval
$\Delta_\alpha$, $\Phi = 0$ is a parametrization fixing condition, and
$\Delta_\Phi$ is the associated determinant. The paths $x(\lambda)$ go
from the endpoint $x^\prime$ to the endpoint $x^{\prime\prime}$. We will not need the
further details of the measure or of the conditions on the paths in $p$
and $N$; they can be found in \cite{Harpp}.

The sum-over-histories 
construction of the decoherence functional parallels (\ref{onesix}) and
utilizes the Klein-Gordon product.  
Initial and final spacelike surfaces $\sigma'$ and $\sigma''$ are
B
selected on which the Klein-Gordon products in (\ref{onesix}) are
evaluated. The analog of the unit operator is supplied by the
sum over all paths in (\ref{fivethree}).

However, there is an important difference between the sum-over-histories
construction just adumbrated and the product space decoherence schemes
described in Subsection C. The matrix elements defined by (\ref{onesix})
do not satisfy the constraints over the whole of configuration space
(as in (\ref{onefour}) for the most general and interesting choices
of the integration. They cannot be taken to be the matrix elements
of a class operator in a product space decoherence scheme. 
A simple argument for understanding this follows.

We may express the characteristic function in (\ref{fivethree}) as a Fourier
transform, creating an exponent containing the effective action
\begin{equation}
S[p,x,N] + \mu F[p,x,N]
\label{fivefour}
\end{equation}
where $\mu$ is the parameter of the Fourier transform. The formal
manipulations which argue that invariantly constructed path 
integrals satisfy the constraints ({\it e.g.}~\cite{HH91}) would, in this
case, lead to the conclusion that the $\langle x^{\prime\prime} || 
\widetilde C_\alpha || x^\prime \rangle$ satisfy an operator version of the
equation $\delta S/\delta N + \mu \delta F/\delta N=0$. This is indeed
the constraint when $F$ is a reparametrization-invariant functional of
$x^A$ and $p_A$ alone --- the kind of ``observable'' usually considered
in canonical quantization. But it is not the case for the more general
reparametrization invariant functionals that can depend on $N(\lambda)$ 
As a result, there is some subtlety in using (\ref{fivethree})
to define a `class operator' of the product space schemes
of Section IIC. 

It is now straightforward to show that generalized sum-over histories
quantum mechanics can be written as a product space decoherence scheme.
By solving the constraint equation using
$\langle x^{\prime\prime} || \widetilde C_\alpha || x^\prime\rangle$  on
$(\sigma'',\sigma')$ and
its normal derivative as initial data
on these surfaces, it should be possible to construct a function $C(x'',x')$ on
$\cal Q \times \cal Q$ which satisfies the
constraints everywhere and whose value and normal derivative coincide
with those of $\langle x^{\prime\prime}|| \widetilde C_\alpha ||
x^\prime\rangle$ on the surfaces ($\sigma^\prime, \sigma^{\prime\prime}$).
The class operators thus defined will depend on the specification of these
surfaces. 
The product space decoherence scheme constructed from the
resulting $C(x'',x')$ and the Klein-Gordon product will then
coincide with that obtained by the
sum-over-histories prescription.

\section{The induced and Klein-Gordon Products}
\label{prelim}

We now review the formulation of the induced product and show
that it can be related to the Klein-Gordon product for
many of the AFH models described in Section II. 
This relationship is a generalization of the 
observation of \cite{Mar94} that the induced product
corresponds to a kind of `absolute
value' of the Klein-Gordon product when the metric $G$ is flat and the
potential vanishes.  This simple relationship will be of
great use in comparing the Klein-Gordon and induced product based
decoherence schemes in Section IV below.
Many readers may find
the relationship between the induced and Klein-Gordon inner products
to be their best source of intuition
when dealing with the induced product.  As a result, some readers
may wish to glance quickly at Section IIIB before studying the review
of the induced product in IIIA.

\subsection{ Review of the induced Product}
\label{sec:IIIA}

We now briefly review the induced product.  This discussion
is intended to provide only a passing familiarity
with the scheme to allow the unfamiliar reader to follow certain
calculations and to have a general understanding of
the results.  For this reason, we consider only the most straightforward
application of the techniques of \cite{ALMMT95,Mar95b,Lan95}
to the asymptotically
free hyperbolic models of Section IIA;
more general and more detailed treatments can be 
found in \cite{Higsum,ALMMT95,Lan95,Mar94,Mar95a,HT83} and especially 
\cite{Mar95b}.

The induced product is motivated by attempts to construct
a physical inner product for Dirac-style canonical quantization
\cite{Dir64} of constrained systems. 
Following Dirac, we will be
interested only in solutions $\psi$ of the constraint equation
$H\psi = 0$.  Such solutions give the so-called `physical
states,' and it is on these states that the induced product will
be defined.

We consider an AFH model as in section IIB and associate with
our system the `auxiliary' Hilbert space ${\cal H}_{\rm aux}
= L^2({\cal M}, \sqrt{-G} d^nx)$; 
this space is called
auxiliary because the final physical states will not live in 
${\cal H}_{\rm aux}$ --- they will not be normalizable in its
inner product.  However, we will use this space to `induce'
an inner product on a space ${\cal H}_{\rm phys}$ of physical
states.  We are interested in the case where $H$ is self-adjoint and has
purely continuous spectrum on ${\cal H}_{\rm phys}$; this
occurs whenever $({\cal Q}, G)$ is asymptotically flat and $V$
decays sufficiently fast in a sufficiently large region near
infinity.  The inner product of two states $|\phi\rangle$ and
$|\psi \rangle$ in ${\cal H}_{aux}$ will be denoted by
$\langle \phi || \psi \rangle$.

In this situation, and under a certain further technical assumption 
concerning the operator $H$, the physical Hilbert space
is not difficult
to construct.  Note that what we would really like is to `project'
${\cal H}_{\rm aux}$ onto the (generalized) states which are
zero-eigenvalue eigenvectors of
$H$.  Of course, since none of these states are normalizable, this
will not be a projection in the technical sense.  Instead, it will
correspond to an object which we will call $\delta(H)$, a Dirac
delta `function.'  Given the above mentioned assumption on
$H$ (see \cite{Mar95b}), the object $\delta(H)$ can be shown to exist
and to be uniquely defined.  Technically speaking however, it 
exists not as an operator in the Hilbert space ${\cal
H}_{aux}$, but as
a map from a dense subspace ${\cal S}$ of ${\cal H}_{aux}$ 
to the (for our
purposes, topological) dual
${\cal S}'$ of ${\cal S}$.  The space ${\cal S}$
may typically be thought of as a Schwarz space; that
is, as the space of
smooth rapidly decreasing functions (`test functions')
on the configuration space.
In this case, ${\cal S}'$ is the usual space of tempered
distributions.
Not surprisingly, this is reminiscent of 
the study of generalized eigenfunctions through Gel'fand's 
spectral theory \cite{GV64} and ${\cal S} \subset {\cal
H}_{aux}
\subset {\cal S}'$ forms a rigged Hilbert triple.

The key point is as follows:  While generalized eigenstates
of $H$ do not lie in ${\cal H}_{aux}$, they 
can be related to normalizable
states through the action of the `operator' $\delta(H)$.  That is, 
generalized eigenstates $|\psi_{\rm phys}\rangle$ of $H$ with 
eigenvalue $0$, can always be expressed in the form $\delta(H)|\psi_0
\rangle$, where $|\psi_0\rangle$ is a normalizable state in
${\cal S} \subset {\cal H}_{aux}$.  This choice
of $|\psi_0\rangle$ is of course not
unique and, in fact, we associate with the physical state
$|\psi_{\rm phys}
\rangle$ the entire {\it equivalence class} of normalizable
states $|\psi\rangle \in {\cal S}$ satisfying
\begin{equation}
\delta (H) |\psi \rangle = |\psi_{\rm phys} \rangle.
\label{A1}
\end{equation}
Each equivalence class of normalizable states will form a {\it
single} state of the physical Hilbert space.

All that is left now is to `induce' the physical inner product from
the auxiliary Hilbert space.  Naively, the 
inner product of two physical states $\phi_{\rm phys}$ and
$\psi_{\rm phys}$ may be written $\langle \phi || \delta(H)
\delta (H) || \psi \rangle$, where $|\phi \rangle$ and $|\psi \rangle$
are normalizable states in the appropriate equivalence classes.
This inner product is clearly divergent, as it contains
$[\delta(H)]^2$.  The resolution is simply to `renormalize' this
inner product by defining the {\it physical} (induced) product 
$\circ_I$ to be
\begin{equation}
\phi^*_{\rm phys} \circ_I \psi_{\rm phys} =  \langle
\phi || \delta(H) || \psi \rangle.
\label{threethree}
\end{equation}
Note that (\ref{threethree}) does not depend on
which particular states
$|\phi \rangle$,
$|\psi \rangle \in {\cal S}$ were chosen to represent
the physical
states $|\phi_{\rm phys}\rangle$ and $|\psi_{\rm phys}\rangle$.  
This construction parallels the 
case of purely discrete spectrum as, if $P_H$ were a projection
onto normalizable zero-eigenvalue eigenstates of $H$, we would have
$[P_H]^2 = P_H$.  Although $\delta(H)$ is not strictly speaking
an operator, taking $|\phi \rangle $ and $|\psi \rangle$
to lie in ${\cal S}$ makes the above inner product well defined.
As a result, $\circ_I$ is a bilinear product
with the reality properties required to build a product space decoherence
scheme.

\subsection{Relating the Induced and Klein-Gordon Products}
\label{sec:IIIB}

We now show that, for a large class of AFH models, the Klein-Gordon and
induced products are connected by a simple relation.  Indeed,
linearity leads us to expect that {\it any} two bilinear product operations 
$\circ^\prime$ and $\circ^{\prime\prime}$ on the same space 
${\cal V}$ should be related by
\begin{equation}
\phi\, \circ^{\prime\prime}\, \psi = \phi\, \circ^\prime\, A\psi
\label{threefour}
\end{equation}
for some linear operator $A$, up to possible subtleties
concerning the domain. In the remainder of  this section  we show
that there is indeed such a relation between the  Klein-Gordon and induced
products for AFH systems and we exhibit the corresponding 
operator $A$.

The connection is most easily found by introducing the usual apparatus
of $\delta$-function normalized states, for example, the eigenfunctions
$\psi_\lambda$ of the constraints such that
\begin{equation}
H\psi_\lambda = \lambda\psi_\lambda,
\label{threefive}
\end{equation}
normalized so that
\begin{equation}
\left\langle\psi_\lambda || \psi_{\lambda^\prime} \right\rangle =
\delta(\lambda - \lambda^\prime).
\label{threesix}
\end{equation}
It follows from (\ref{threesix}) that if $\psi_\lambda$ is a continuous one
parameter family of eigenfunctions approaching 
$\psi_{\rm phys}$ as $\lambda\to 0$, then
\begin{equation}
\left\langle\psi_\lambda || \phi_{ \rm phys} \right\rangle = \delta
(\lambda) \left(\psi_{ \rm phys}^*\, \circ_I\, \phi_{ \rm phys}\right)
\label{threeseven}
\end{equation}
This is what was meant by (\ref{oneten}).  But, for our systems, 
the left hand side of
(\ref{threeseven}) may be evaluated in terms of the Klein-Gordon product
on a spacelike slice $\sigma$
yielding the connection between $\circ_{KG}$ and the induced product.
We will always restrict to the case $\lambda \leq 0$ so that
the Klein-Gordon product is in fact independent of $\sigma$, so long
as $\sigma$ ends only at spatial infinity. 

There are a number of particular cases according to how the potential
behaves at large positive times.  In considering these, it is useful to
keep in mind that, with the sign conventions of (\ref{twoone}), it is
$-V$ that would function like an effective potential as far as one
dimensional motion in $t=x^0$ is concerned.  The simplest case is when
there is a single asymptotic free region at $t\to -\infty$ and $V(x)\to
-\infty$ sufficiently fast 
at $t\to + \infty$ so that all generalized eigenstates of $H$
must vanish there\footnote{In order for a generalized
eigenstate to lie in ${\cal S}'$ as required, it must be a
tempered distribution.  That is, it may increase as $x^0 \rightarrow
+\infty$, but not in an exponential manner.}.
Effectively, there is a repulsive barrier for
propagation to large, positive, time. We begin with this situation.

The explicit form of the inner product on the left hand side of
(\ref{threeseven}) is
\begin{equation}
\left\langle\psi_\lambda || \phi_{\rm phys} \right\rangle = \int d^nx
\psi^*_\lambda (x) \phi_{\rm phys} (x).
\label{threeeight}
\end{equation}
The key point is that this integral must yield the $\delta$-function form
on the right hand side of (\ref{threeseven}) with no finite additions.
Any finite range of the integral in $t$ will not contribute to the
singular $\delta$-function. Neither will the asymptotic region at $t\to
+ \infty$ since we have assumed that the wave functions vanish there.
Only the asymptotic region $t\to-\infty$ contributes to
the coefficient of the $\delta$-function.  There, $\psi_\lambda$ may be
expanded in terms of a complete set of positive and negative frequency
solutions of $\nabla^2\psi = \lambda \psi$ in the form
\begin{equation}
\psi_\lambda(x) = \int d^{n-1} p \left[a^{(+)}_{\lambda\vec p}
f^{(+)}_{\lambda\vec p} (x) + a^{(-)}_{\lambda \vec p}
f^{(-)}_{\lambda\vec p} (x)\right]
\label{threenine}
\end{equation}
for some coefficients $a^{(+)}_{\lambda\vec p}$, $a^{(-)}_{\lambda\vec
p}$. Explicitly, {\it e.g.}
\begin{equation}
f^{(+)}_{\lambda\vec p}(x) = \left[(2\pi)^3 (2\omega_{\lambda
p})\right]^{-\frac{1}{2}} \exp \left[i\left(-\omega_{\lambda p} t + \vec p
\cdot \vec x\right)\right]
\label{threeten}
\end{equation}
for the $3+1$ case, where
\begin{equation}
\omega_{\lambda p} = \left(\vec p^{\,2} - \lambda\right)^{\frac{1}{2}}.
\label{threeeleven}
\end{equation}
Since  $\lambda \leq 0$ and, in this region, the metric is flat and the
potential is a nonnegative constant,
 the separation of positive and negative frequency
states is well defined.  This provides a definition of the `positive
and negative frequency parts' $\phi^{(+)}_\lambda$, $\phi^{(-)}_\lambda$
of $\phi_\lambda$.
There is a similar expression for $\phi_{\rm phys}(x)$.

The integral (\ref{threeeight}) in the
asymptotic region $t\to -\infty$ may be evaluated to
yield the coefficient of $\delta(\lambda)$ in (\ref{threeseven}) (and
hence the induced product) in terms of the constants $a^{(+)}_{\lambda \vec
p}$, $a^{(-)}_{\lambda\vec p}$ and the similar constants for $\phi_{\rm
phys}$. Not surprisingly, since only an asymptotic regime of $t$ is
involved, this expression can also be evaluated in terms of the
Klein-Gordon product on a surface of constant $t$. The following
relation between the induced and  Klein-Gordon products emerges when the
potential exhibits a single asymptotically free region 
\begin{equation}
\left(\psi_{\rm phys}\, \circ_I\, \phi_{\rm phys}\right) = \pi
\left[\left(\psi^{(+)}_{\rm phys}\, \circ_{KG}\, \phi^{(+)}_{\rm phys}\right)
- \left(\psi^{(-)}_{\rm phys}\, \circ_{KG}
\, \phi^{(-)}_{\rm phys}\right)\right].
\label{threetwelve}
\end{equation}

\noindent It is not necessary to indicate on which surface the Klein-Gordon
product is evaluated because, being between solutions of the
constraints for $\lambda \le 0$ and $V \ge 0$ at infinity,
it is independent of the surface (so long
as this surface ends only at spatial infinity).  The important point is
that the decomposition of $\phi_{\rm phys}$ and $\psi_{\rm phys}$ into
their positive and negative frequency parts occurs at the
asymptotic surface $t \rightarrow - \infty$.  In this respect
induced product methods are similar to the approach advocated by
Wald in \cite{W}.

The relation between the induced and Klein-Gordon products
may be expressed more compactly by introducing the operator
\begin{equation}
\Omega^{(-)} = \pi {\rm sign} \left(p^{(-)}_0\right)
\label{threethirteen}
\end{equation}
where $p^{(-)}_A$ is the momentum in the asymptotic region $t\to -
\infty$. Then we have
\begin{equation}
\psi_{\rm phys}\, \circ_I\, \phi_{\rm phys} = \psi_{\rm phys}\, \circ_{KG}
\, A \phi_{\rm phys}
\label{threefourteen}
\end{equation}
with $A=\Omega^{(-)}$.  It is important to note that when
$\Omega^{(-)}$ is interpreted as an operator from states normalizable
in the induced product to states normalizable in the Klein-Gordon
product, it is not surjective (even on a dense subspace).  In fact, 
any solution $\psi$ that is normalizable in the induced product
decreases rapidly at $x^0 \rightarrow + \infty$ so that its
Klein-Gordon norm vanishes.  The Klein-Gordon norm of $\Omega^{(-)}
\psi$ then vanishes as well.  As a result, the above relation can
be inverted to give the Klein-Gordon product in terms
of the induced product  only for a special set of states
despite the fact that an inverse for $\pi {\rm sign} \left(p^{(-)}_0\right)$
exists on the space of Klein-Gordon normalizable states.

Other asymptotic forms of the potential give the relation (\ref{threefourteen})
between
the induced and Klein-Gordon products but with differing operators $A$.
For example, for systems that are asymptotically free in both past
and future (as with the
free relativistic world
line), a form of (\ref{threefour}) holds with
\begin{eqnarray}
A&=& \Omega^{(+)} + \Omega^{(-)}\\
 &=& \pi \left[{\rm sign} \left(p^{(+)}_0\right) + {\rm sign}
\left(p^{(-)}_0\right)\right] \equiv \Omega.
\label{threefifteen}
\end{eqnarray}
In this case the relation (\ref{threefourteen}) {\it is} invertible. 
Thus we may also write
\begin{equation}
\label{tworeg}
\psi_{\rm phys} \circ_{KG} \phi_{\rm phys} = \psi_{\rm phys}
\circ_I \Omega^{-1} \phi_{\rm phys}.
\end{equation}

The expression (\ref{threefifteen}) can be used to write
the inner product in terms of the `Bogoliubov coefficients'
associated with the constraint equation $H \psi = 0$.
For example, suppose that two solutions $\psi$ and
$\psi'$ are purely positive frequency in the far past, but
that $\psi = \alpha + \beta$ and $\psi' = \alpha' + \beta'$
where $\alpha,\alpha'$ are purely positive frequency in the
far future and $\beta, \beta'$ are purely negative frequency
in the far future.  Then we have
\begin{equation}
\psi^*\circ_I \psi' = 2 \pi \alpha^* \circ_{KG} \alpha'.
\end{equation}

Another interesting case is provided by a potential of the
form  (\ref{twofive}) with $\Lambda > 0$.
The potential $V$ then  approaches $-\infty$ at
$t\to+\infty$, becoming infinitely attractive for motion in $t$. Indeed, it
is sufficiently attractive that the constraint $H$ is not automatically
self-adjoint but permits various self-adjoint extensions. These are
clearly discussed in \cite{CFGM90}. The various self-adjoint extensions
are equivalent to inserting an impenetrable ``wall'' at some large value
of $t$. As far as the relation between the induced and Klein-Gordon
products this case is therefore the same as $V\to -\infty$. Thus we have
(\ref{threefourteen}) with $A=\Omega^{(-)}$.

We expect that our asymptotically free assumption may be relaxed
somewhat and that (\ref{threetwelve}) or (\ref{tworeg})
will continue to hold.  In support of this conjecture, note that
when the $t$-dynamics is separable (so that it decouples from the other
($\vec{x}$) coordinates) and the Klein-Gordon norm is conserved, these
results can be derived without imposing additional
assumptions on the dynamics of the $\vec{x}$ coordinates.
As a result, we would 
not be surprised to find that (\ref{threetwelve}) holds also for complicated
models such as Bianchi IX cosmologies.

\section{Klein-Gordon and Induced product space schemes}

The connections between the Klein-Gordon and induced products uncovered
in the previous Section naturally give rise to the question of whether
there are corresponding relationships between the product space
decoherence schemes which employ them. We shall investigate such
relationships in this Section, especially to see when two different
decoherence schemes yield equivalent physical predictions. 
We will consider specifically the case of AFH models with two asymptotically
free regions and the case of a single asymptotically free region in 
the past and a sufficiently repulsive potential in the future.
When the spaces of states are appropriately chosen and the other
structures are chosen in the usual way, 
product space decoherence schemes based on the Klein-Gordon and induced
products are isomorphic in the sense of section IIA.  However, 
this always involves choosing a space of states which is smaller
than what one would naively use for one of the schemes. 
That is not necessarily a disadvantage in quantum cosmology
where one deals with fixed initial and final conditions prescribed
by fundamental laws. 

Let us begin with a schematic description of a notion of equivalence between
product space decoherence schemes which is a restriction of the
notion of isomorphism between decoherence schemes described in
Section IIA. For the moment we ignore 
issues such as invertibility of operators, domains, and so on.
We will then state below how the spaces of states may be chosen so that
our schematic argument does in fact define an isomorphism of decoherence
schemes.  Below, the 
involution $*$ is complex conjugation of functions on the
configuration space ${\cal Q}$.

The results of Section IIIB tell us that a change of product
(from Klein-Gordon to induced) can always be compensated by a vector space
isomorphism $A$ on the space of states.  That is, for a fixed class operator
$C$:
\begin{equation}
\phi^* \circ_{KG} C \circ_{KG} \psi = (A \phi)^* \circ_I C \circ_I
(A \psi)
\end{equation}
where we have used the fact that the operator $A$ from Section IIIB
is always real under $*$ and
self-adjoint with respect to $\circ_I$.  Let us take
the partitions $\Pi$ to be defined only by the condition that
$\sum_\alpha C_\alpha = U$.  Then the constraint H and the 
the unit alternatives $U_{KG}$ and
$U_{I}$ (to be used with the Klein-Gordon and induced schemes
respectively) provide the only remaining general structures in a product
space decoherence scheme. The constraint $H$ will be preserved
if $H$ commutes with $A$. If the
operators $\hat{U}_I = U_I\circ_I $ and $\hat{U}_{KG} = U_{KG} \circ_I$ 
are invertible, then the pair 
\begin{equation}
\label{iso}
\sigma = A,  \ \    
\chi = \hat{U}_{I}\hat{U}^{-1}_{KG} \otimes \openone
\end{equation}
is an isomorphism from the Klein-Gordon decoherence scheme to the
induced scheme that preserves $H$.

The operators $A$ that relate the products have already been considered
in Section IIIB. Since these operators involve only the signs of
asymptotic momenta, they commute with the constraint $H$ and solutions
of the constraint are mapped into solutions of the constraint by $\sigma$.
It remains to discuss 
the unit alternatives $U_{KG}$ and $U_I$ on whose invertibility
the isomorphism depends.  We take as our model for
Klein-Gordon schemes
the sum-over-histories schemes of \cite{Harpp}, in which the
unit alternative $U_{KG}$ is based on the Feynman propagator $\Delta_F$,
which is in turn defined by the path integral of the form (\ref{fivethree})
with a sum over positive lapse and all paths. 
As remarked in Section IIC, the result is not a class operator
of the kind
needed for a product space scheme as $\Delta_F$
does not fully satisfy the constraints.  (Indeed, $\Delta_F$ is well
known to be a Green's function for the constraint $H$.)  However, 
following Section IIC, we can translate this unit operator into 
the language of product
space schemes.  The result is that this unit alternative becomes
just the `positive
frequency function' $U_{KG} = G^{(+)}$,
the bi-solution to the constraints that is
positive frequency in $y$ when $y$ is far to the future and is
negative frequency in $x$ when $x$ is far to the past.  
The appendix shows that, for the case of two asymptotically free
regions, the matrix elements of $G^{(+)}$ are just
\begin{equation}
\label{GME}
G^{(+)}(x,y) = \langle x || \Pi^+_{+ \infty} \delta(H) \Omega ||y \rangle
\end{equation}
where $\Pi^+_{+\infty}$ ($\Pi^+_{-\infty}$)
again denotes the projection onto solutions
of the constraint that are purely positive frequency in the far future
(past).  Now $\Omega$ is invertible (in fact, its inverse is
bounded on the space of induced normalizable states) but, due to the
presence of the projections in (\ref{GME}), the invertibility of $G^{(+)}$
will depend on choosing the proper space of states.

In the case where only the past is asymptotically free (and where the
potential is sufficiently repulsive in the future) the
appendix shows that,
\begin{equation}
G^{(+)}(x,y) = -2 \pi \langle x||\Pi^+_{-\infty} \delta(H)||y\rangle.
\end{equation}
Again, any invertibility issues will rest on the
proper choice of state space.

On the other
hand, in schemes based on the induced product, it is natural to choose
the unit alternative to be represented by the matrix elements
$U_I(x'',x') = \langle x''||\delta(H)||x'\rangle$.
This is because $\delta(H) \circ_I \psi$ is just $\psi$ itself.
The constructions of \cite{Mar95b,Mar94,Mar95a,BDT} may
be thought of as providing a
decoherence scheme of this type.  Since $\delta(H) \circ_I$ is just
the unit operator on ${\circ}_I$-normalizable states it is in general
invertible.

We now provide the appropriate details to turn our schematic
argument above into an actual isomorphism.  We begin with the
Klein-Gordon scheme for the case of two asymptotically flat regions.
Because of the projections inherent in $G^{(+)}$, we must restrict
ourselves to, say, states  which are positive frequency in the
far future (else the normalization coefficient ${\cal N}$  (\ref{onesix b})
would diverge).  In particular, let us take the space ${\cal V}_N^{(KG)}$
on which the Klein-Gordon product is to be defined to consist
of those smooth functions on ${\cal Q}$ which are purely positive
frequency in the far future and are normalizable in ${\cal H}_I$.
Note that $\Omega^{-1}: {\cal V}_N^{(KG)} \rightarrow \Omega^{-1}
{\cal V}_N^{(KG)}$
(defined by  (\ref{tworeg})) is injective.  As a result, if we define the 
induced product on the space ${\cal V}_N^{(I)} \equiv \Omega^{-1}
{\cal V}_N^{(KG)}$, the invertibility of the vector space map
($\Omega^{-1}$) is assured.  Our schematic argument will then
provide an isomorphism so long as $\hat{U}_{KG}^{-1}$ can be defined as a map
from ${\cal V}_N^{(I)} = \Omega^{-1} {\cal V}_N^{(KG)}$ to the space
of $\circ_I$ normalizable states.  Now, for $\phi \in {\cal V}_N^{(KG)}$,
we have
\begin{equation}
\Omega^{-1} \hat{U}_{KG} (\Omega ^{-1} \phi) = \Omega^{-1} \phi
\end{equation}
since $\phi$ is positive frequency in the far future.  As a result, we
may take $\hat{U}_{KG}^{-1} = \Omega^{-1}$ and define an isomorphism
of decoherence schemes through (\ref{iso}).
Note, however, that the space of states chosen for the induced
scheme is considerably smaller than the Hilbert space used in, for example, 
\cite{Mar94}.

For the case of a single asymptotically flat region (and such that the
potential is sufficiently repulsive in the far future), it is convenient to
consider the induced product scheme first.  We will take the space 
${\cal V}_N^{(I)}$ on which $\circ_I$ is defined to consist of those smooth
solutions to the constraints with finite norm in ${\cal H}_I$.  The
Klein-Gordon scheme will use ${\cal V} = \Omega^{(-)} {\cal V}_N^{(I)}$.
In this case, the required isomorphism is just
$(\sigma = \Omega^{(-)}, \chi = \openone \otimes \openone)$.  To show
this, we simply note that if $\psi = \Omega^{(-)}\alpha$,
$\phi = \Omega^{(-)} \beta$ for $\alpha, \beta \in {\cal V}_N^{(I)}$
then
\begin{equation}
\psi^* \circ_{KG} U_{KG} \circ_{KG} \phi
= - \pi \alpha^* \circ_I \beta
\end{equation}
since
$(\alpha^{(+)})^* \circ_{KG} \beta^{(+)} = - 
(\alpha^{(-)})^* \circ_{KG} \beta^{(-)}$, where $\alpha^{(\pm)},
\beta^{(\pm)}$ are the positive and negative frequency parts of
$\alpha,\beta$ (in the distant past).
The constant $-\pi$ is irrelevant in computing the decoherence 
functional and again
we see that the relation between the products has lead directly to
an isomorphism of decoherence schemes.    However, this time it is the
space of states for the Klein-Gordon scheme which has to be artificially
restricted.

\section{Conclusion}

Generalized quantum theory is a comprehensive 
framework for quantum theories of closed systems.  As a result, its
principles can be implemented in many ways that are different
from the way they are implemented in the usual quantum theory.
Classical physics, for example, can be
considered as a generalized quantum theory \cite{Har91a}. This generality 
may be needed to
deal with dynamical quantum spacetime geometry.

However, as the present paper illustrates, there is less freedom in the
construction of generalized quantum theories than might naively be
supposed. Two decoherence schemes may be equivalent in the
sense that they yield identical decoherence functionals for corresponding
sets of alternative histories.  The predictions for the probabilities of
decoherent sets of alternative histories are then isomorphic.  This
occurs when the elements representing
histories, boundary conditions, and auxiliary structures can be mapped
into one another preserving all of the associated decoherence functionals.

In this paper we have compared
certain decoherence schemes appropriate for the quantum
mechanics of systems, such as
the relativistic world line and homogeneous minisuperspace
models, which have a single reparametrization invariance.
Each scheme involves a
bilinear product $\circ$, but there will often be an
isomorphism between decoherence schemes constructed from two different
products, or perhaps between appropriate restrictions of 
such schemes.  We have seen that schemes based on 
the Klein-Gordon and induced products are equivalent
(when appropriately restricted) in models  with either one or two
asymptotically free regions.  Note that, due to the restriction of the
state space, it is natural to restrict the set of class operators
(or observables) as well.

Equivalences of the kind described here will be useful in
narrowing the choice of decoherence functionals for
reparametrization-invariant systems and, in addition, 
they may also allow the utilization
of different techniques for calculation and approximation corresponding
to the different but equivalent ways in which
the decoherence functional can be
expressed.

\acknowledgements
This work was supported in part by NSF grants  PHY94-07065 and PHY94-07194.
DM would also like to thank the Erwin Schr\"odinger Institute for
their hospitality during work on this project and for their partial 
support.  DM also thanks Syracuse University for their partial
support and Jos\'e Mour\~ao for useful discussions.

\appendix

\section{The class operator $U = G^{(+)}$ }

In this appendix, we will derive a number of useful expressions
for the class operator $G^{(+)}(x'',x')$ in various AFH models.
Recall that this class operator is the bi-solution of the constraint
constructed from the Feynman propagator $\Delta_F(x,y)$ for
$x$ near past infinity and $y$ near future infinity.

We begin by considering models with two asymptotically free
regions, one to the past and one to the future. 
For such models, the spaces of states that are normalizable
in the induced and Klein-Gordon 
products are essentially the same.  In addition, for such cases
our system resembles a scattering experiment; the states may be
thought of as free in both the distant past and the far future.
Here, it is important to keep in mind that `past' and `future'
as we use the words have no direct physical meaning in terms of the
scattering experiment, but merely label regions of the configuration
space.  The most direct analogy is to, for example, a one
dimensional quantum mechanical scattering problem in which
the states are free both on the far left and on the far right.

In this appendix, we will make use of the space ${\cal H}_{aux} 
= L^2({\bf R}^n)$
and adapt our notation accordingly.  The Feynman propagator can
be defined as the operator 
\begin{equation}
\label{fpop}
\Delta_F = -i\int_0^{\infty} dN \exp(iNH) = {1  \over {H + i \epsilon}}
\end{equation}
on this space.
It is convenient to write the matrix elements of
$\Delta_F$ in terms of a complete set of eigenfunctions $\psi^+_{k,
}$
of $H$.  Here we choose $k=(k_0, \vec{k})$ to label the asymptotic
momenta in the far future; this is indicated by the superscript
$+$ on the  wave functions.    The eigenfunctions satisfy
\begin{equation}
H \psi^+_{k} 
= (-k_0^2 + \vec{k}^2 + (m^+)^2)\psi^+_{k} 
= (k^2 + (m^+)^2)\psi^+_{k_0,\vec{k}} 
\end{equation}
where $(m^+)^2$ is the asymptotic value of the potential in the far
future.  
We will take these states to be normalized so as
to have the inner products
\begin{equation}
\langle \psi_{k_0,\vec{k}}^+ || \psi_{p_0,\vec{p}}^+
 \rangle = \delta^n(k-p).
\end{equation}
As a result, the Feynman propagator takes the familiar form
\begin{equation}
\label{efrep}
\langle x || \Delta_F || y \rangle = \int  d^n k {{\psi^{+}_{k} (x)
\psi^{+*}_k(y)} \over {k^2 + (m^+)^2 + i \epsilon}}.
\end{equation}

We wish to compute the Klein-Gordon product (on the right) of this expression
with a solution $\psi(y)$ to the constraint equation.  This product
can be evaluated on any hypersurface $\Sigma$ lying to the future
of the point $x$ and having its boundary at spatial infinity.
  Let us therefore take this surface to lie in the
far future so that $\psi_{k}(y)$ becomes just an oscillatory 
exponential with wave vector $k$.  For $y \gg |x|$, 
we may perform the $k_0$
integral in (\ref{efrep}) by the usual method of closing the contour
in the upper half plane.  The contour encloses only the pole
at positive frequency, so that we obtain  
\begin{equation}
\label{solrep}
\langle x || \Delta_F || y \rangle = 2 \pi i \int  d^{n-1} \vec{k}
 {{\psi^{+}_{\omega(\vec{k}),\vec{k}} (x)
\psi^{+*}_{\omega(\vec{k}),\vec{k}} (y)}
\over {2\omega(\vec{k})}} 
\end{equation}
where $\omega(\vec{k}) = \sqrt{\vec{k}^2 + (m^+)^2}$. 
It is then clear that the Klein-Gordon product of $\Delta_F$
and $\psi$ yields just $i$ times the part of $\psi$ which is
positive frequency in the far future. Denoting the associated
projection by $\Pi^+_{+\infty}$, we have
\begin{equation}
G^{(+)}\circ_{KG} \psi = - \Pi^+_{+\infty} \psi,
\end{equation}
where the Klein-Gordon product is taken on a surface in the far
future of the point at which both sides are evaluated.

Similarly, $\psi^* \circ_{KG} G^{(+)} = - (\Pi^+_{-\infty} \psi)^*$
with the product evaluated at $x^0 \rightarrow - \infty$.
Taking the
Klein-Gordon product on both sides of $\Delta_F$ (as
$x^0 \rightarrow \pm \infty$ respectively) and using the fact that
the Klein-Gordon and induced products are related by a factor
of $\Omega$ allows one to derive the relation
$\Pi^+_{+ \infty} \delta(H) \Omega = \Omega \delta(H) \Pi^+_{-\infty}$. 
The relation
can also be checked by more direct means.
Note that the matrix elements of the class operator $U = G^{(+)}$
may therefore be written 
\begin{equation}
\label{matel}
U(x,y) = \langle x || \Pi^+_{+\infty} \delta(H) \Omega || y \rangle. 
\end{equation}

We now turn to the case of a single asymptotic region as
in Section IIA; we will
take the far past to be asymptotically free.  We proceed much as
before, keeping in mind that we are only interested
in $\Delta_F(x,y)$ for large negative $x$ and large positive $y$.
The matrix 
elements of $\Delta_F$ can again be written in terms of the
eigenfunctions of $H$ on ${\cal H}_{aux}$.  Of course, 
because states that are purely positive or negative frequency in
the far past grow
exponentially in the far future, any solution to the constraint which is
normalizable in the induced product contains both positive and
negative frequency parts in the far past.  
We may, howevever, consider a basis of states $\psi_{k_0,\vec{k}}$ for $k_0 >0$
with asymptotic momentum $\vec{k}$ in the spacelike directions
and which satisfy $P_0^2 \psi_{k}
= k_0^2 \psi_{k}$, where $P_0$ is the asymptotic
momentum (in the far past)
in the $x^0$ direction.  We may then express the Feynman
propagator as
\begin{eqnarray}
\langle x || \Delta_F || y \rangle &=& \int_{k_0 >0 } d^nk 
{{\psi_{k} (x)
\psi^{*}_k(y)} \over {k^2 + (m^-)^2 + i \epsilon}}
\cr
&=& 
\int_{k_0 } d^nk 
{{\phi_{k} (x)
\psi^{*}_k(y)} \over {k^2 + (m^-)^2 + i \epsilon}}
\end{eqnarray}
where $(m^-)^2$ is the asymptotic value of the potential and
$\phi_k(y)$ for $k_0>0$ ($k_0 <0$) is the solution of the
constraint that matches the positive (negative) frequency part of
$\psi_{|k_0|,\vec{k}}$ in the far past; i.e, 
$\phi_{k_0,\vec{k}} + \phi_{-k_0,\vec{k}} = 
\psi_k$.  
For $x \ll -|y|$, we can again close the $k_0$
contour in the upper half plane.  This time, we find
\begin{equation}
\label{onebdy}
\langle x || \Delta_F || y \rangle = 2 \pi i \int  d^{n-1} \vec{k} \Bigl(
{{\phi_{\omega(\vec{k}),\vec{k}} (x)
 \psi^*_{\omega(\vec{k}),\vec{k}}
 (y)}
\over {2\omega(\vec{k})}} \Bigr) 
\end{equation}
where $\omega(\vec{k}) = \sqrt{ \vec{k}^2 + (m^-)^2}$.

Suppose now that $\phi$ is some $\circ_I$-normalizable state.
Due to the form of (\ref{onebdy}), 
we do not expect $\phi \circ_I G^{(+)}$ to have any
well-defined meaning.  However, taking the product with the state $\phi$
on the right yields
\begin{equation}
G^{(+)} \circ_I \phi = -2\pi  \Pi^+_{-\infty} \phi
\end{equation}
where $\Pi^+_{-\infty}$ again denotes the `positive frequency projection'
defined at $t = - \infty$, although $\Pi^+_{-\infty} \phi$ is not
normalizable with respect to the induced product.

It is also interesting to consider the Klein-Gordon product of (\ref{onebdy})
with a $\circ_I$ normalizable state.
Now a solution $\psi_k$ is orthogonal in the Klein-Gordon product
to all $\phi_p$ unless $\vec{p} = \vec{k}$ and
$p_0 = \pm k_0$.  As a result, taking the Klein-Gordon product
of a state $\psi$ (that is normalizable in the induced product) on the
left with
$G^{(+)}$ yields just $\psi \circ_{KG} G^{(+)} = - \sqrt{2}
\psi$.  On the other hand, taking the Klein-Gordon inner product
with a state $\phi$ (which is normalizable in the induced product) on
the right yields $G^{(+)} \circ_{KG} \phi =0$.  This may be
seen from the fact that the inner product may be taken on any
hypersurface which ends only at spatial infinity.  In particular, we
may take this hypersurface to lie in the far future.  There, however, 
both $\phi$ and $\psi_{\omega(\vec{k}),\vec{k}}$ vanish.

\end{document}